\documentclass[twocolumn,preprintnumbers,amsmath,amssymb,superscriptaddress,10pt]{revtex4}

\usepackage{graphicx}
\usepackage{dcolumn}
\usepackage{bm}

\begin{document}


\title{Relationship between the Kramers-Kronig relations and negative index of refraction }

\author{Mark C. Hickey}\altaffiliation[Electronic mail: ]{mark\_hickey@uml.edu}
\affiliation{%
Center for Electromagnetic Materials and Optical Systems, Department of Electrical and Computer Engineering, University of Massachusetts, Lowell, Massachusetts
01854 USA.
}%

\author{Alkim Akyurtlu}
\affiliation{%
Center for Electromagnetic Materials and Optical Systems, Department of Electrical and Computer Engineering, University of Massachusetts, Lowell, Massachusetts
01854 USA.
}%
\author{Adil-Gerai Kussow}
\affiliation{%
Department of Physics, University of Massachusetts, Lowell, Massachusetts
01854 USA.
}%

\date{\today}

\pacs{78.20.Ci, 78.20.Ek, 78.20.Ls and 78.67.Pt}
\keywords{}

\begin{abstract}
The condition for a negative index of refraction with respect to the vacuum index is
established in terms of permittivity and permeability susceptibilities. It is found that the imposition of analyticity to satisfy the Kramers-Kronig relations
is a sufficiently general criterion for a physical negative index. The satisfaction of the  Kramers-Kronig relations is a manifestation of the principle of causality and the predicted frequency region of negative index agrees with the Depine-Lakhtakia condition for the phase velocity being anti-directed to the Poynting vector, although the conditions presented here do not assume {\it a priori} a negative solution branch for n.
\end{abstract}
\maketitle

\begin{figure}[!ht]
\begin{center}
 \includegraphics[width=2.75in]{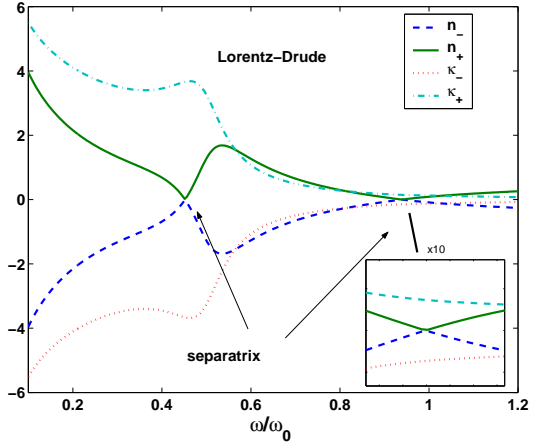}\\
\end{center}
\caption{(Color Online) Plot of the family of solutions for n for which a Drude model is used for $\epsilon$ and a Lorentz model is used for $\mu$.}
\label{Fig1}
\end{figure}

\begin{figure}[!ht]
\begin{center}
 \includegraphics[width=2.75in]{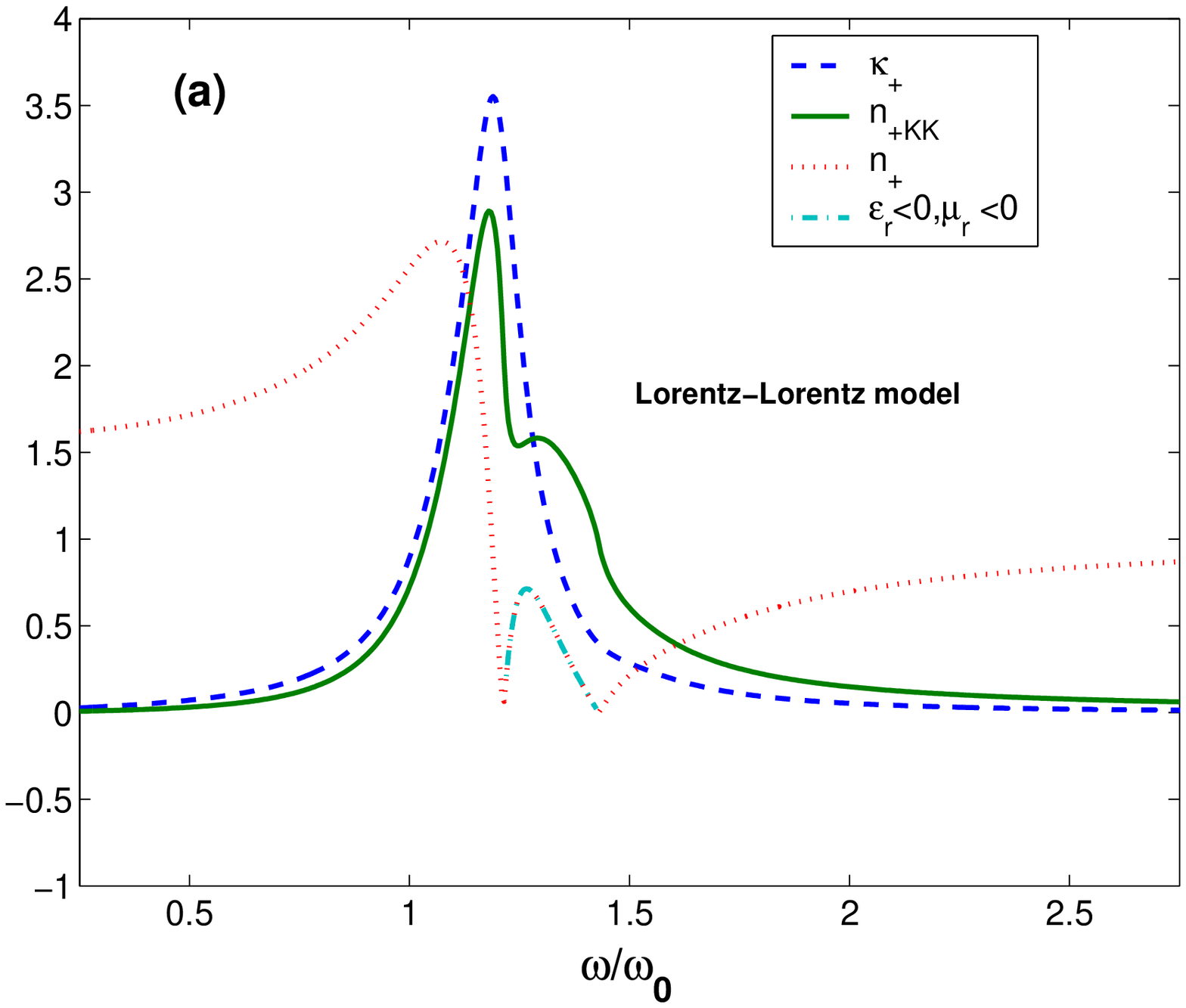}\\
 \includegraphics[width=2.75in]{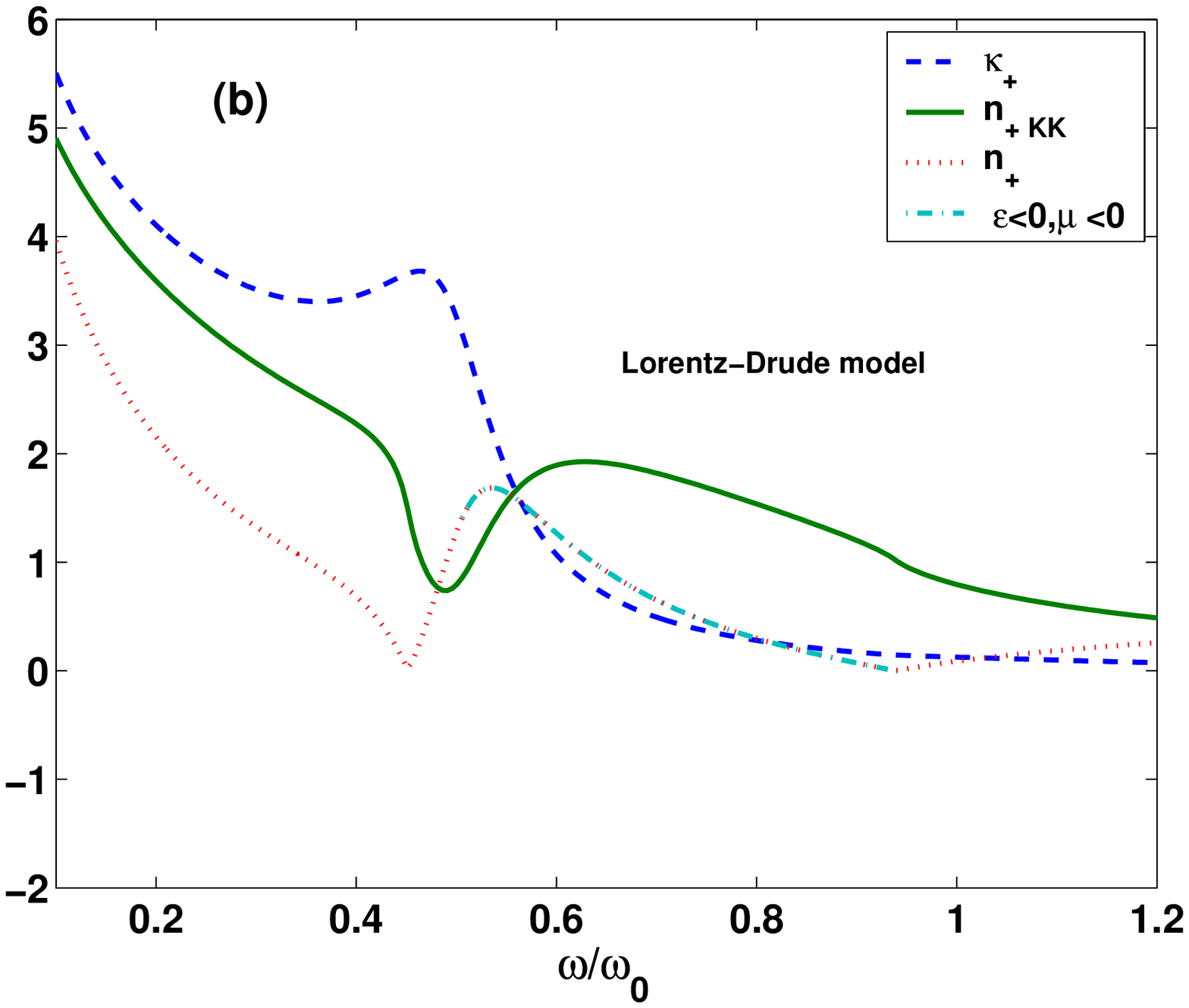}\\
\end{center}
\caption{(Color Online) Plot of the n$_{+}$ solution branch, together with its Kramers-Kronig transformation for the Lorentz-Lorentz model (a) and the Drude-Lorentz model (b). Note that the Kramers-Kronig transformed n does not match $\kappa$.}
\label{Fig2}
\end{figure}

Negative index optics is an attractive field because it is driven by a wealth of exotic applications such as the creation of a perfect
perfect lens \cite{PhysRevLett.85.3966}, cloaking \cite{J.B.Pendry05252006} and through-wall vision \cite{PhysRevLett.102.253902}. While much success has been made of actualizing a negative refractive index in `meta-materials', there are still some theoretical loose ends in the understanding of
the physical conditions in which negative refraction can occur, quite apart from considerations of efficiency losses, material choice and experimental proof of principle.
Experiments by Shelby \cite{R.A.Shelby04062001} and Smith \cite{PhysRevLett.85.2933}
 have shown negative refraction in the microwave region of the electromagnetic spectrum using meta-materials, while metal/dielectric hybrid rod structures allowed
 Yao \cite{JieYao08152008} {\it et al.} to demonstrate negative refraction in the visible
 spectral region.\\
  The experimental proof of negative refraction lies in the arena of geometric ray bending in slabs and prisms as well as being derivable from the transmittance and reflectance data.
The original criteria which were postulated for a material to possess a negative index of refraction (n=Re[N])
were given by a paper by Veselago \cite{Sov.Phys.Usp.10.509} in 1968, wherein he established that, in order to have a negative index, the real parts of $\epsilon$ and $\mu$ need to be simultaneously negative. In this scenario, the magnetic induction, electrostatic displacement and the wave-vector form a left-handed triad and the material possesses a negative index in the frequency region wherein the Veselago condition is satisfied. However, it has emerged that the Veselago conditions are not sufficiently general to capture all eventualities of negative n, because they are only valid for a purely real permittivity and permeability. An important paper by Depine and Lakhtakia \cite{Micro.Opt.Techn.Lett.41.315} showed that, having assumed the negative branch of the solution for n, the
condition of having the phase velocity anti-directed to the group velocity (and Poynting vector) gives a more general criterion for negative refraction. This was a generalization of the concept of negative refraction to
negative phase velocity. Nevertheless, there exists disagreement in the literature as exemplified by the theoretical presentation of a case by Valanju {\it et al.} \cite{PhysRevLett.88.187401} of
positively refracting waves whose group and phase velocities were non-collinear in an inhomogeneously dispersed wave packet in a negative index material.\\
McCall \cite{McCall200892} considered a space-time covariant formalism of the calculation of phase velocity  in negative index materials. Mackay and Lakhtakia \cite{PhysRevB.79.235121} recently considered the effect of bianisotropy
in determining the directionality dependence of negative index, phase and group velocity conditions in terms of the complex wave-vector for pseudo-chiral
omega materials. In this work, we consider scalar responses, but we examine the
  model independent criteria for negative index in terms of the response functions. Stockman \cite{PhysRevLett.98.177404} considered the square of the complex refractive index N$^{2}$ and applied the
Kramers-Kronig (KK) relations (if $\epsilon$ and $\mu$ are causal response functions, N$^{2}$ is analytic and also obeys the KK relations) to relate the phase and group velocities to the dissipation in the system. This approach worked with complex N$^{2}$ instead of the real n and $\kappa$.
Since the latter two are the optical constants which determine the phase velocity and group velocity, and plane wave attenuation, it is the purpose of this letter to work with these and impose the KK relations at this level.
Kinsler \cite{PhysRevA.79.023839} finds dealing with the refractive index under the definition n=$\sqrt{Re(N^{2})}$ more convenient because one does not need to make corrections for spatial oscillations of the counter propagating electric field at interfaces. Nevertheless, there is still a sign choice implicit in this definition and he appeals to Dephine and Lakhtakia's work \cite{Micro.Opt.Techn.Lett.41.315} for the criterion for negative n which $\epsilon$ and $\mu$ must satisfy. Peiponen \cite{Eur.Phys.J.B.41.61} showed that, having solved for the refractive index using the Veselago conditions, the real and complex parts of the refractive index satisfied the Kramers-Kronig relations. The purpose of this letter is
to show that the satisfaction of the Kramers-Kronig relations should be the primary criterion for determining the correct sign of n.\\
The above considerations suggest a need of a more robustly general way of looking at the conditions for negative n with respect to the vacuum refractive index.
In this paper, we wish to establish the most general criteria for negative n (Re[N]), by examining the solution branches for N$^{2}$=$\epsilon\mu$, and choosing the solution in each frequency region which maintains analyticity (differentiability along the real axis of n).  The choice of the solution such that the
functions n($\omega$) and $\kappa$($\omega$) (the extinction) are analytic implies that these functions, being  the real and complex parts of a function N=n+i$\kappa$, must satisfy the Kramers-Kronig relations \cite{PhysRev.104.1760}. This is an indirect manifestation of causality at the level of electromagnetic plane waves - there can be no dispersion without absorption.\\
Let us examine all possible solutions to N$^2$=($\epsilon\mu$), where the permittivity and permeability are given by $\epsilon$ and $\mu$. Writing $\epsilon$=$\epsilon_{r}$+i $\epsilon_{i}$ and using a similar expression for $\mu$, we arrive at the following quartic :
\begin{align}
n^4-\left(\frac{\epsilon_{r}\mu_{i}+\epsilon_{i}\mu_{r}}{2}\right)^{2}-(\epsilon_{r}\mu_{r}-\epsilon_{i}\mu_{i})n^{2}=0.
\end{align}
 A similar equation exists for $\kappa$. There are four solutions for n, two of which are purely imaginary and two of which are real and these two are labeled n$_{+}$ and n$_{-}$. The two real solutions are written as follows :
\begin{align}
n_{\pm}=\pm\frac{\sqrt{\epsilon_{r}\mu_{r}-\epsilon_{i}\mu_{i}+|\epsilon||\mu|}}{\sqrt{2}}
\end{align}
These solution branches are plotted in Figure \ref{Fig1}, and it can be seen that there
are two separatrices. For this plot, we have used one model whereby a Lorentzian resonance is used for both the permittivity and permeability (Lorentz-Lorentz), while the second model assumes a Drude plasmon model for the permittivity response and a Lorentzian for the permeability (Drude-Lorentz model). The models are described in the following way :
\begin{align}
\epsilon_{D}(\omega)=\epsilon_{\infty}\left(1-\frac{\omega_{pe}^2}{\omega(\omega-i\gamma_{e}\omega_{pe})}\right)\\
\epsilon_{L}(\omega)=1+\frac{\omega_{pe}^2}{\omega_{0e}^2-\omega^{2}+i\gamma_{e}\omega}\\
\mu(\omega) =1+\frac{\omega_{pm}^{2}}{\omega_{0m}^{2}-\omega^{2}+i\gamma_{m}\omega},
\end{align}
where the parameters used for the model calculation are chosen to be $\omega_{0m}$=0.5, $\omega_{0e}$=1.15, $\omega_{pe}$=1.0, $\omega_{pm}$=0.8, $\gamma_{e}$=0.3, $\gamma_{m}$=0.11,  $\epsilon_\infty$=0.4 and a frequency scale
of $\omega_{0}$=1.\\
The locations of the separatrix in Fig \ref{Fig1} are two points in frequency space at which the curves for n$_{\pm}$ are no longer differentiable (analytic). We can see this clearly by calculating
$\partial n_{+}/\partial \omega$ and this is written as follows :
\begin{align*}
\frac{\partial n_{+}}{\partial \omega} = \frac{1}{\sqrt{2}}\frac{1}{2}\left(\frac{\epsilon_{r}'\mu_{r}+\epsilon_{r}\mu_{r}'-(\epsilon_{i}'\mu_{i}+\epsilon_{i}\mu_{i}')
}{{\sqrt{\epsilon_{r}\mu_{r}-\epsilon_{i}\mu_{i}+|\epsilon||\mu|}}}\right)\\
+\frac{1}{\sqrt{2}}\frac{1}{2}\left(\frac{|\epsilon|^{'}|\mu|+|\epsilon||\mu|^{'}}{\sqrt{\epsilon_{r}\mu_{r}-\epsilon_{i}\mu_{i}+|\epsilon||\mu|}}\right),
 \end{align*}
where $\mu^{'}$ indicates, for example, differentiation with
respect to $\omega$.
As can be seen from Figure \ref{Fig2}, the pairs of solutions (n$_{-}$,$\kappa_{-}$) and (n$_{+}$,$\kappa_{+}$) do not satisfy the Kramers-Kronig relations. This is because, as can be seen from the equation for $\partial n_{+}/\partial \omega$ above , the solution branch n$_{+}$ departs from analyticity (by possessing a divergent derivative) at the frequency which satisfies the following equation :
\begin{equation}
\epsilon_{r}\mu_{r}-\epsilon_{i}\mu_{i}+|\epsilon||\mu|=0.
\label{condition1}
\end{equation}
Mathematically, n$_{+}$($\omega$) and n$_{-}$($\omega$) are not analytic because of presence of the absolute value function.\\
 The solutions to Equation \ref{condition1} define the locations of the separatrix in Fig \ref{Fig1}. The departure of analyticity here means that, without the correct sign choice, the group refractive index would be divergent and un-physical. Noting that, $|\epsilon|=\sqrt{\epsilon_{r}^{2}+\epsilon_{i}^2}$, we can simplify Equation \ref{condition1} to read :
\begin{equation}
\epsilon_{r}=-\frac{\epsilon_{i}\mu_{r}}{\mu_{i}}.
\label{condition2}
\end{equation}
This equation admits at least two solutions $\omega_{\pm}$ for the Lorentz-Lorentz and Drude-Lorentz models used here, and the number of solutions in general depends on the model for $\epsilon$ and $\mu$ used. The condition above in Equation \ref{condition2} is actually identical to that which was suggested by Ruppin \cite{R.Ruppin} in a book chapter
 \cite{neg.phase.book} and previous article \cite{Micro.Opt.Techn.Lett.47.313}. The condition is equivalent
 to the Depine-Lakhtakia condition \cite{Micro.Opt.Techn.Lett.41.315} except for a sign ambiguity.
  If we choose the solution for n such that
\begin{align*}
                    &n_{+}(\omega),\;   0 < \omega <\omega_{-}\\
n(\omega)=\big\{   &n_{-}(\omega),\;  \omega_{-} < \omega <\omega_{+}\\
                    &n_{+}(\omega),\;  \omega_{+} < \omega <\infty
\end{align*}
which now satisfies the Kramers-Kronig relation \cite{Lucarini} :
\begin{align*}
\kappa(\omega)=-\frac{2}{\pi}\displaystyle\int_{0}^{\infty}\frac{(n(\omega^{'})-1)\omega^{'}}{\omega^{'2}-\omega^{2}}d\omega^{'},
\end{align*}
which can be seen from Figure \ref{Fig4}. This scheme for choosing the sign of the refractive
index requires the frequency range to be calculated by Equation \ref{condition2}, and the phase angle
of N$^{2}$ (=(n+i$\kappa$)$^{2}$) lies in the within the interval [-$\pi$,0]
in the frequency range $\omega_{-} < \omega <\omega_{+}$, while it lies in [0,$\pi$] everywhere else.
For the Lorentz-Lorentz case computed here give $\omega_{-}$=1.21 and $\omega_{+}$=1.43 in units of $\omega_{0}$.\\
 \begin{figure}[!ht]
\begin{center}
 \includegraphics[width=2.75in]{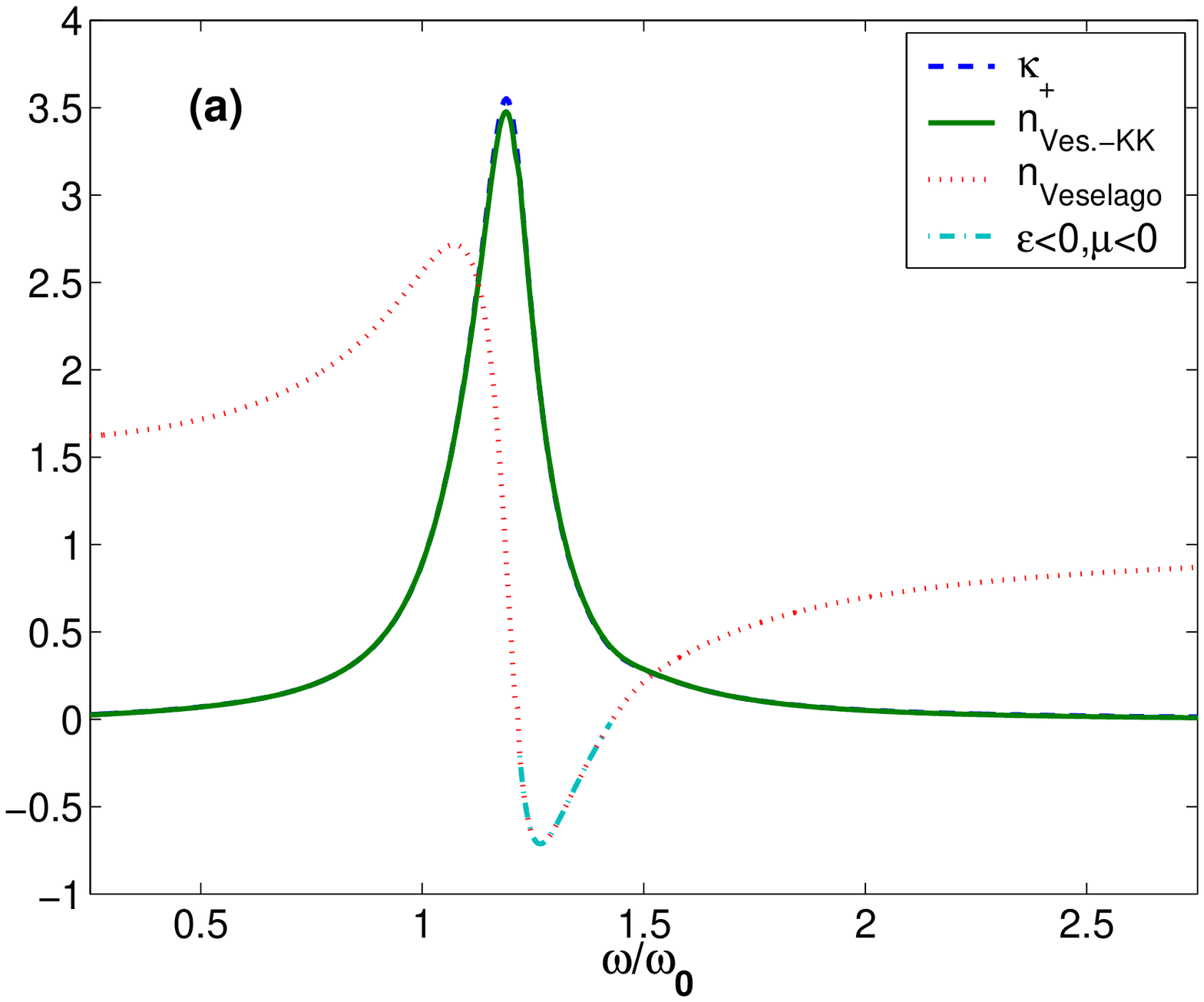}\\
 \includegraphics[width=2.75in]{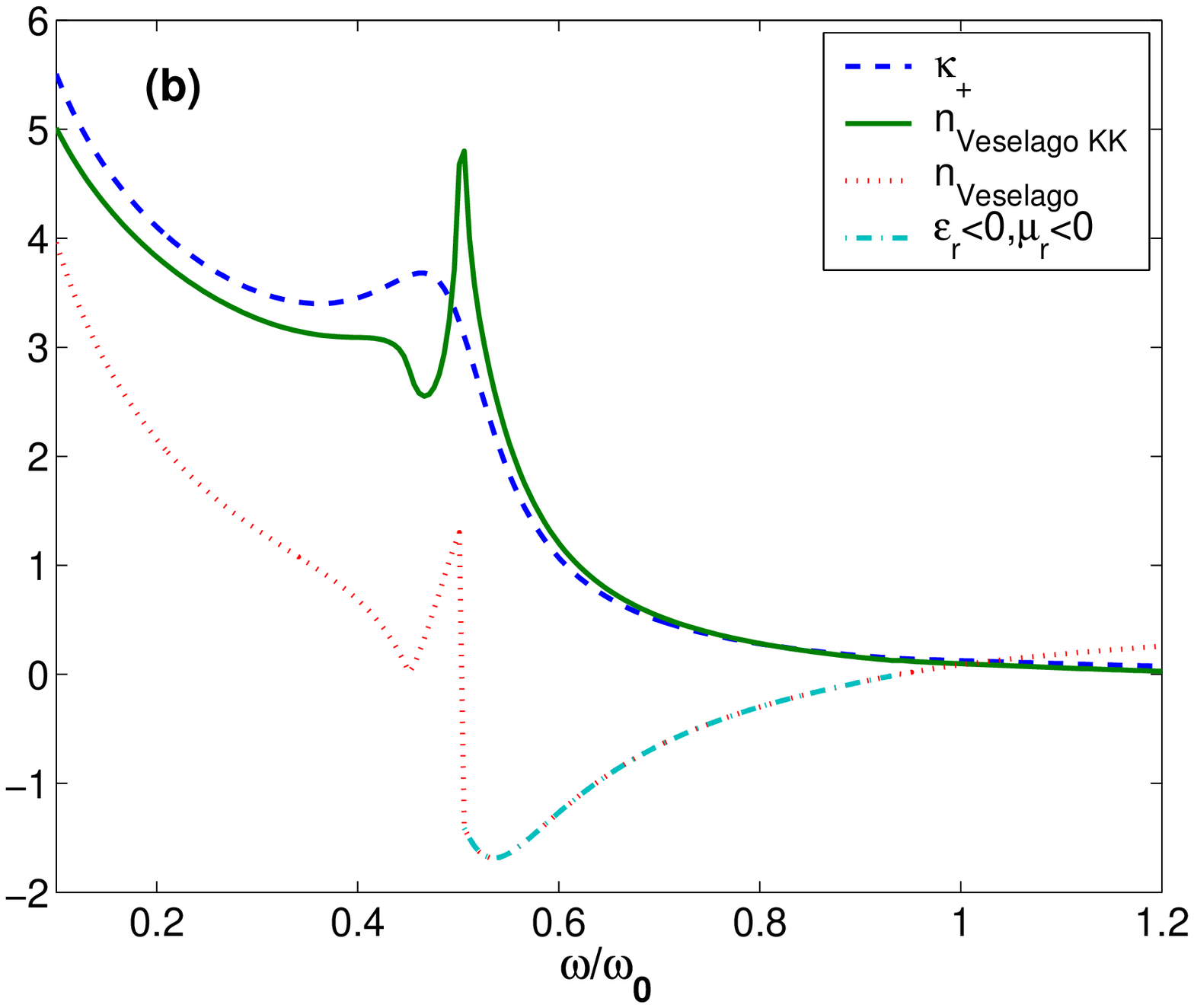}\\
\end{center}
\caption{(Color Online) (a) Plot of the refractive index under the Veselago condition (choosing the n$<$0 solution branch (n$_{-}$) when $\epsilon$ and $\mu$ are simultaneously negative) for the Lorentz-Lorentz model. (b) a comparison between that Veselago refractive index solution and the solution determined by the Depine-Lakhtakia condition. We see that
that Veselago criterion is not sufficiently general, because it does not sustain the
validity of the Kramers-Kronig relations.}
\label{Fig3}
\end{figure}
 \begin{figure}[!ht]
\begin{center}
 \includegraphics[width=2.75in]{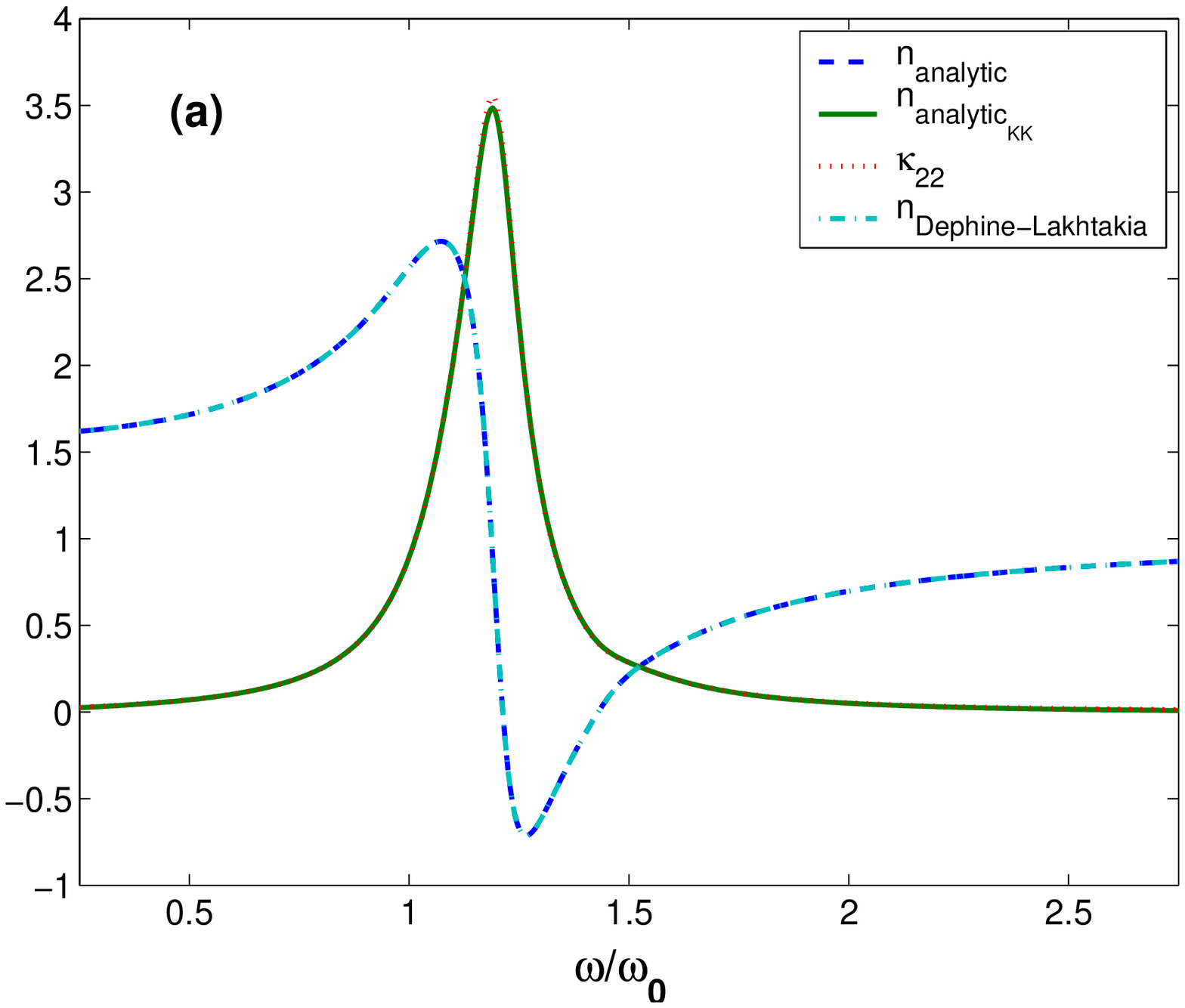}\\
 \includegraphics[width=2.75in]{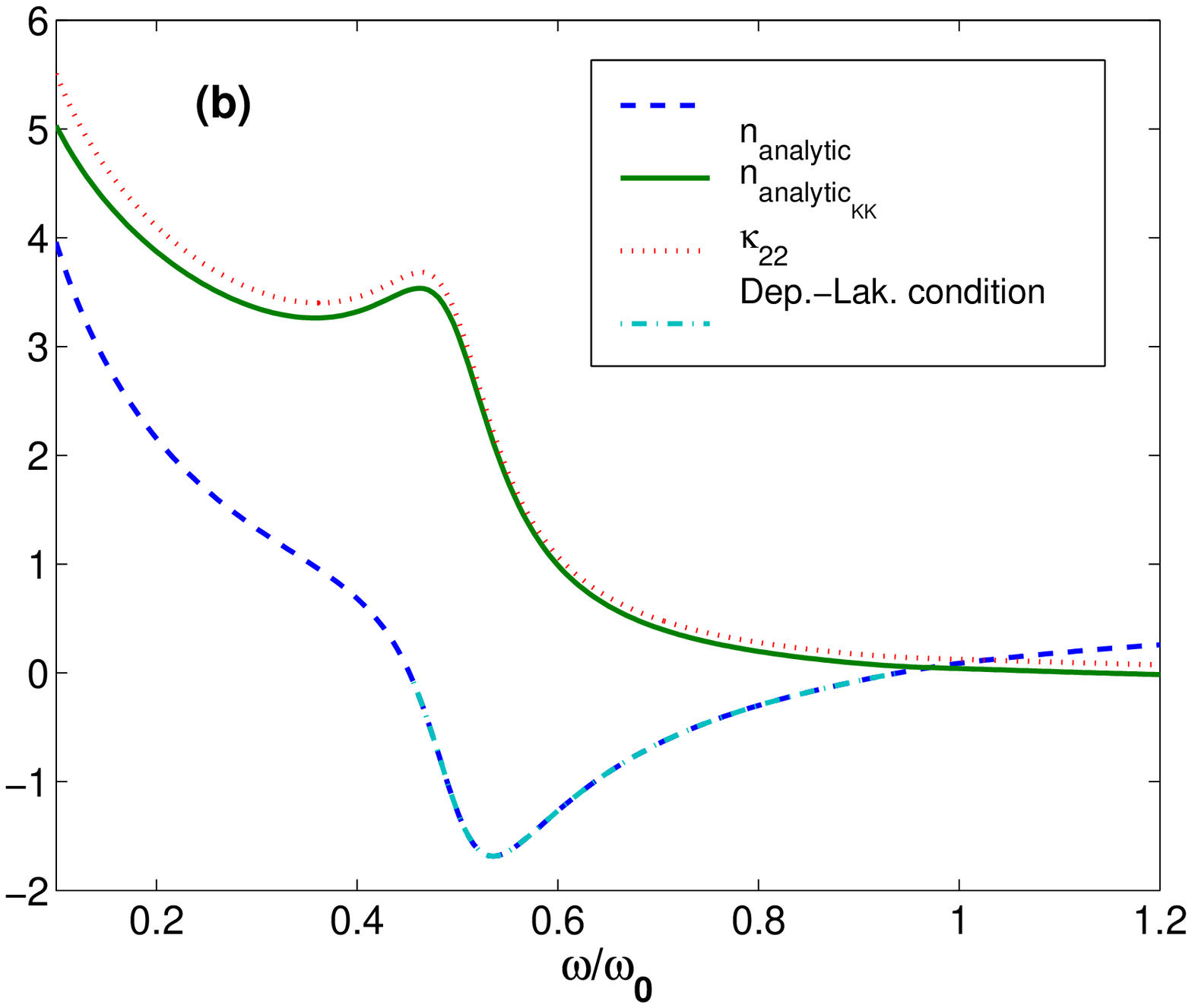}\\
\end{center}
\caption{(Color Online) (a) a comparison between the Depine-Lakhtakia condition and the analytic condition for the Lorentz-Lorentz model. (b) The analytic condition imposed on the solution branch restores the Kramers-Kronig relations , and hence causality. }
\label{Fig4}
\end{figure}
These are the frequency regions for which we must switch to the negative branch n$_{-}$, in order to maintain differentiability in n($\omega$) . This now defines a set of (n($\omega$),$\kappa(\omega$)) which satisfies the Kramers-Kronig (KK) relations, as plotted in Fig \ref{Fig4} (a). This requirement of analyticity to satisfy the KK relations is a manifestation of the relationship between the real and complex parts of an analytic function and is a manifestation of the principle of causality. It is shown in Figure \ref{Fig3} that the Veselago conditions are not sufficiently general to capture the frequency interval over which the refractive index is negative. If we were to choose the n$_{-}$ only in the region where Re($\epsilon$) and Re($\mu$) were both less than zero, we would violate analyticity.\\
It can be seen from Figure \ref{Fig4} that the predicted frequency region of negative n agrees with
those given by the Depine-Lakhtakia condition \cite{Micro.Opt.Techn.Lett.41.315}, which
was established from considering that the phase velocity vector should be anti-directed to the Poynting vector. This comes as a corollary to our Kramers-Kronig criteria here, most simply encapsulated in Equation \ref{condition2}.\\
 We have found that the principle of causality alone, as expressed in the Kramers-Kronig relations, is sufficient to establish whether a material possesses a negative index of refraction with respect to vacuum, and if there is a negative index, the frequency range for which the refractive index is negative are given. In frequency space, causality implies that the refractive index is analytic in the upper-half complex plane, unless the responses of the system are active, wherein the condition of analyticity is sustained by choosing the integration contour for the Kramers-Kronig relations such that poles lie on a line below this contour, as described by Skaar \cite{PhysRevE.73.026605}. The analyticity condition is consistent with those of the Dephine-Laktakia analysis, but we don't not need to assume {\it a priori} that n $<$0. In this sense, the
choice of the negative branch based on sustaining analyticity of n($\omega$) and $\kappa$($\omega$)
(and hence to satisfy of the Kramers-Kronig relations) is the most general criterion.
\begin{acknowledgments}
M. C. Hickey is grateful to the Trinity for the uniformity of nature. We wish to thank Anas Mokhlis for fruitful discussions and Shawn Belmore for reading the manuscript.
\end{acknowledgments}


\end{document}